\documentclass[showpacs,prd]{revtex4}
\usepackage{epsfig}
\usepackage{graphicx}%
\usepackage{dcolumn}
\usepackage{amsmath}
\usepackage{latexsym}

\begin{document}

\title{Operator equations and Moyal products -- metrics in quasi-hermitian quantum mechanics}
\author{F G Scholtz} 
\altaffiliation{fgs@sun.ac.za}
\author{H B Geyer} 
\altaffiliation{hbg@sun.ac.za}
\affiliation{Institute of Theoretical Physics, University of
Stellenbosch,\\ Stellenbosch 7600, South Africa}
\date{\today}

\begin{abstract}
The Moyal product is used to cast the equation for the metric of a non-hermitian Hamiltonian in the form of 
a differential equation.  For Hamiltonians of the form $p^2+V(ix)$ with $V$ polynomial this is an exact equation.  
Solving this equation in perturbation theory recovers known results.  Explicit criteria for the hermiticity and 
positive definiteness of the metric are formulated on the functional level. 

\end{abstract}

\maketitle

\section{Introduction}
\label{intro}

The recent interest in PT-symmetric quantum mechanics \cite{bbj,bender} has revitalized the old question 
\cite{scholtz,most1,most2} of the existence of a metric and associated inner product for which a standard quantum 
mechanical interpretation is possible, even though the Hamiltonian may be non-hermitian with respect to the 
given inner product. Here we address this issue with technology borrowed from non-commutative quantum 
mechanics.  The advantage of this approach is that the operator equation that must be solved can 
often be cast in the form of a differential equation without making any approximations. Generically 
this equation may be of infinite order, but in many cases of physical interest it turns out to be finite. 
This equation contains all the information required to construct the metric operator exactly.  In addition 
criteria can be formulated to test the hermiticity and positive definitness of the metric directly on the 
level of this equation, leading to considerable simplification.  On this level the non-uniqueness of 
the metric is reflected in the choice of boundary conditions.  On the other hand it is known \cite{scholtz} 
that the metric is uniquely determined (up to an irrelevant normalization factor) once a complete set of 
irreducible observables has been specified which is hermitian with respect to the inner product associated 
with the metric.  This suggests an interplay between boundary conditions in phase space and the choice of 
physical observables. 

\section{A Moyal product primer}
\label{Moyal}
\subsection{Finite dimensional Hilbert space}
\label{finite}

Although the Moyal product is a well established tool \cite{moyal}, recently revived in the context of 
non-commutative systems (see e.g. \cite{fairlie}), we review the construction briefly in order to 
adapt it to our specific application. We start by considering the construction of an irreducible unitary 
representation of the Heisenberg-Weyl algebra
\begin{equation}
\label{algebra}
gh=e^{i\phi}hg;\quad g^\dagger=g^{-1},\, h^\dagger=h^{-1}
\end{equation}
on a finite dimensional Hilbert space with dimension $N$. Clearly, such a representation can only exist 
for non-trivial $\phi$ if ${\rm tr}\; gh=0$.  The implication of this is clear when we compute the trace explicitly.  
Since $g$ is unitary, 
\begin{equation}
\label{eigeng}
g|\alpha\rangle=e^{i\alpha}|\alpha\rangle,\,\alpha\in R\,;\quad 
\langle\alpha^\prime|\alpha\rangle=\delta_{\alpha^\prime,\alpha}\,.
\end{equation}
From (\ref{algebra}) it follows that
\begin{equation}
\label{ladder}
gh|\alpha\rangle=e^{i(\alpha+\phi)}h|\alpha\rangle,
\end{equation}
so that $h$ ladders between the eigenvalues of $g$.  We conclude that the eigenvalues and eigenstates of $g$ are 
of the form
\begin{eqnarray}
\label{eigensysg}
|\alpha_0+n\phi\rangle&=&h^n|\alpha_0\rangle\equiv |n\rangle,\\
g|\alpha_0+n\phi\rangle&=&e^{i(\alpha_0+n\phi)}|\alpha_0+n\phi\rangle\nonumber,
\end{eqnarray}
with $\alpha_0$ an arbitrary constant (set to zero without loss of generality). If the 
representation is to be irreducible, all $N$ orthogonal eigenstates of $g$ can be reached through 
such a laddering process.  

It is now simple to compute ${\rm tr}\; gh=0$.  The result is
\begin{equation}
{\rm tr}\; gh=\sum_{n=0}^N e^{in\phi}=\frac{1-e^{iN\phi}}{1-e^{i\phi}},
\end{equation}
which vanishes only when $\phi=\frac{2m\pi}{N}$ with $m$ integer.  This limits the allowed values of $\phi$. 
Note from (\ref{eigensysg}) that $g^N|n\rangle=|n\rangle$, $h^N|n\rangle=|n\rangle$, $\forall n$, implying 
$g^N=h^N=1$. The operators $g^n$ and $h^m$ are therefore only independent when $n,m<N$. Motivated by this, 
we choose $\phi=\frac{2\pi}{N}$ as with this choice the operators $U(n,m)\equiv g^nh^m$, with 
$n=0,1,\ldots N-1$ and $m=0,1,\ldots N-1$, form a basis in the space of operators (matrices) on the 
Hilbert space.   To show the linear independence and completeness of this basis we introduce the standard 
inner product on the space of operators
\begin{equation}
\label{trnorm}
(A,B)={\rm tr}\; A^\dagger B.
\end{equation}
It immediately follows from (\ref{eigensysg}) that $(U(n^\prime, m^\prime),U(n,m))
=N\delta_{n^\prime,n}\delta{m^\prime,m}$, implying that these operators are linearly independent. 
As there are $N^2$ such complex linearly independent operators, it follows on simple dimensional 
grounds that they provide a basis. 

The conditions above are necessary for the existence of a representation of (\ref{algebra}), but 
we have not yet demonstrated that such a representation actually exists.  This follows from
explicit construction.  It is easy to verify that the following matrices satisfy all the conditions above 
\cite{fairlie1}
\begin{eqnarray}
\label{matrices}
g_{n,m}&=&e^{\frac{2\pi i(n-1)}{N}}\delta_{n,m}\,\nonumber\\ 
h_{n,m}&=&\delta_{n,m-1}+\delta_{n,N}\delta_{m,1}\,.
\end{eqnarray}
As the operators $U(n,m)\equiv g^nh^m$ form a basis, any operator $A$ can be expanded in the form
\begin{equation}
\label{expandf}
A=\sum_{n,m=0}^{N-1} a_{n,m}g^nh^m\,,\quad a_{n,m}=(U(n,m),A)/N\,.
\end{equation}
Consider the multiplication of two operators $A$ and $B$
\begin{equation}
\label{multiplyf}
AB=\sum_{n,m=0}^{N-1}\sum_{n^\prime,m^\prime=0}^{N-1} a_{n,m}b_{n^\prime,m^\prime} 
e^{-i m n^\prime \phi}g^{n+n^\prime}h^{m+m^\prime}\,.
\end{equation}
Apart from the phase $e^{-i m n^\prime \phi}$ this looks like the multiplication of two sums in 
which $g$ and $h$ are treated as ordinary complex numbers.  One may therefore take the point of 
view that $g$ and $h$ are to be treated as complex numbers, but then the product rule must be 
modified to ensure equivalence with (\ref{multiplyf}). Making the following substitutions
\begin{equation}
g\rightarrow e^{i\alpha}\,\quad h\rightarrow e^{i\beta}\;,\alpha\,,\beta\in [0,2\pi)
\end{equation}
in the expansion (\ref{expandf}), turns $A$ into a function $A(\alpha,\beta)$, uniquely determined by the 
operator $A$
\begin{equation}
\label{functionf}
A=\sum_{n,m=0}^{N-1} a_{n,m}e^{in\alpha}e^{im\beta}\,.
\end{equation}
To establish an isomorphism with the product (\ref{multiplyf}) we define the Moyal product of functions 
$A(\alpha,\beta)$ and $B(\alpha,\beta)$ \cite{moyal,fairlie}
\begin{equation}
\label{moyal}
A(\alpha,\beta)\ast B(\alpha,\beta)=A(\alpha,\beta)e^{i\phi\stackrel{\leftarrow}{\partial_\beta}
\stackrel{\rightarrow}{\partial_\alpha}}B(\alpha,\beta)\,,
\end{equation}
where the notation $\stackrel{\leftarrow}{\partial}$ and $\stackrel{\rightarrow}{\partial}$ denotes that 
the derivatives act to the left and right, respectively.  On this level operators are replaced by functions, 
as described by (\ref{functionf}), while the non-commutative nature of the operators is captured by the Moyal 
product.  It is easily checked that the Moyal product is associative, as one would expect from the associativity 
of the corresponding operator product.  Once the function $A(\alpha,\beta)$ is given, the coefficients $a_{n,m}$ 
are computed from a simple Fourier transform.  Insertion of these coefficients in (\ref{expandf}) enables the 
reconstruction of the operator.   On a technical point, if one wants to preserve the feature $g^N=h^N=1$, the 
values of $\alpha$ and $\beta$ should actually be restricted to $2m\pi/N$ with $m$ integer.  It is, however, 
simple to take this into account by simply evaluating the function $A(\alpha,\beta)$ only at these values after 
all computations have been performed.

Finally, we derive the relation between the functions $A^\dagger(\alpha,\beta)$ and $A(\alpha,\beta)$ 
corresponding to the hermitian conjugate operator $A^\dagger$ and the operator $A$, respectively.  
Introducing the convention $a_{N+n,N+m}=a_{n,m}$, one easily finds from (\ref{expandf}) that the expansion 
of $A^\dagger$ reads
\begin{equation}
\label{expandfdag}
A^\dagger=\sum_{n,m=0}^{N-1} a^\dagger_{n,m}g^nh^m\,,\quad a^\dagger_{n,m}=a^\ast_{-n,-m}e^{-imn\phi}\,.
\end{equation}
On the level of the functions $A^\dagger(\alpha,\beta)$ and $A(\alpha,\beta)$ this implies the relation
\begin{equation}
\label{hc}
A^\dagger(\alpha,\beta)=\sum_{n,m=0}^{N-1} a^\dagger_{n,m}e^{in\alpha}e^{im\beta}=\sum_{n,m=0}^{N-1} 
a^\ast_{-n,-m}e^{-imn\phi} e^{in\alpha}e^{im\beta}=e^{i\phi\partial_\alpha \partial_\beta}A^\ast(\alpha,\beta)\,.
\end{equation}
An operator is then hermitian if and only if
\begin{equation}
\label{ccf}
A^\ast(\alpha,\beta)=e^{-i\phi \partial_\alpha\partial_\beta}A(\alpha,\beta)\,.
\end{equation}

\subsection{Quantum mechanics}

Here we generalize the results of the previous section to the case of an infinite dimensional quantum system.  
We consider for simplicity the case of one particle in one dimension as the generalization to many particles and 
higher dimensions is obvious. 

In this case a well known irreducible unitary representation of the Heisenberg-Weyl algebra exists \cite{bratteli}
\begin{equation}
\label{weyl}
e^{it\hat p}e^{is\hat x}=e^{i\hbar ts}e^{is\hat x}e^{it\hat p}\,,
\end{equation}
where $\hat x$ and $\hat p$ are the hermitian position and momentum operators satisfying canonical commutation 
relations.  Introducing the notation $U(t,s)\equiv e^{it\hat p}e^{is\hat x}$ and the inner product (\ref{trnorm}), 
one easily verifies
\begin{equation}
(U(t^\prime,s^\prime),U(t,s))=\frac{2\pi}{\hbar}\delta(s-s^\prime)\delta(t-t^\prime)\,,
\end{equation}
which implies as before that these operators are linearly independent.  As before these operators constitute 
a complete set \cite{bratteli} and any operator can be expanded as 
\begin{equation}
\label{expandq}
A(\hat x,\hat p)=\int_{-\infty}^{\infty} dsdt\; a(t,s) e^{it\hat p}e^{is\hat x}\,,\quad a(t,s)=\frac{\hbar}{2\pi} 
(U(t,s),A)\,.
\end{equation}
Note that this expansion reflects the fact that any operator can, due to the irreducibility of the set  $\hat x$ 
and $\hat p$, be written as a function of $\hat x$ and $\hat p$.  Forming the product of two operators 
$A(\hat x,\hat p)$ and $B(\hat x,\hat p)$ one has
\begin{equation}
\label{multiplyq}
A(\hat x,\hat p)B(\hat x,\hat p)=\int_{\infty}^{\infty} dsdt ds^\prime dt^\prime\; a(t,s)b(t^\prime,s^\prime) 
e^{-i\hbar st^\prime} e^{i(t+t^\prime)\hat p}e^{i(s+s^\prime)\hat x}\,.
\end{equation}
Apart from the phase $e^{-i\hbar st^\prime}$ this product looks, as before, like the product of two integrals 
in which $\hat x$ and $\hat p$ are treated as real numbers.  Taking this point of view we can, as before, 
replace $\hat x$ and $\hat p$ by real numbers which turns $A(\hat x,\hat p)$ into a function $A(x,p)$, 
uniquely determined by $A$
\begin{equation}
\label{functionq}
A(x,p)=\int_{-\infty}^{\infty} dsdt\; a(t,s) e^{itp}e^{isx}\,.
\end{equation}
To maintain the isomorphism with the product (\ref{multiplyq}) we have to introduce the Moyal product of 
these functions
\begin{equation}
\label{moyalq}
A(x,p)\ast B(x,p)=A(x,p)e^{i\hbar \stackrel{\leftarrow}{\partial_x}\stackrel{\rightarrow}{\partial_p}}B(x,p)\,.
\end{equation}
On this level we again work with functions, rather than operators, while the non-commutativity of the operators 
is captured by the Moyal product.  As before associativity is easily verified.  Once the function $A(x,p)$ has 
been determined, the function $a(t,s)$ is determined from a Fourier transform.  Insertion into the expansion 
(\ref{expandq}) recovers the operator $A(\hat x,\hat p)$.  

Finally, we derive the relation between the functions $A^\dagger(x,p)$ and $A(x,p)$ corresponding to the hermitian 
conjugate operator $A^\dagger(\hat x,\hat p)$ and the operator $A(\hat x,\hat p)$, respectively.  
From (\ref{expandq}) we easily find that the expansion of $A^\dagger(\hat x,\hat p)$ reads
\begin{equation}
\label{expandqdag}
A^\dagger(\hat x,\hat p)=\int_{-\infty}^{\infty} dsdt a^\dagger(t,s) e^{it\hat p}e^{is\hat x}\,,\quad 
a^\dagger(t,s)=a^\ast(-t,-s)e^{-i\hbar ts}\,.
\end{equation}
On the level of the functions $A^\dagger (x,p)$ and $A(x,p)$ this implies the relation
\begin{equation}
\label{hcq}
A^\dagger(x,p)=\int_{-\infty}^{\infty} dsdt a^\dagger(t,s) e^{itp}e^{isx}=\int_{-\infty}^{\infty} dsdt a^\ast(-t,-s) 
e^{-i\hbar ts}e^{itp}e^{isx}=e^{i\hbar \partial_x\partial_p}A^\ast(x,p)\,.
\end{equation}
This implies that an operator is hermitian if and only if
\begin{equation}
\label{ccq}
A^\ast(x,p)=e^{-i\hbar \partial_x\partial_p}A(x,p)\,.
\end{equation}

In what follows we shall often encounter situations where the operator $A$ is a function of $\hat x$ or $\hat p$ 
only.  It is therefore worthwhile to consider this situation briefly.  Consider the case where $A(\hat p)$. 
From (\ref{expandq}) we have
\begin{equation}
\label{ponly}
a(t,s)=\frac{\hbar}{2\pi} (U(t,s),A(\hat p))=\frac{\hbar}{2\pi}tr(e^{-is\hat x}e^{-it\hat p}A(\hat p))
=\delta(s)\int \frac{dp}{2\pi} A(p)e^{-itp}\,.
\end{equation}
Substituting this result in (\ref{functionq}) we note that the function $A(x,p)$ corresponding to the operator 
$A(\hat p)$ is just $A(p)$, i.e., we just replace the momentum operator by a real number. Clearly, the same 
argument applies to $A(\hat x)$.   

An approach related to the one we discuss here was developed in \cite{bender1}, although in that case the 
position and momentum operators are used as a basis to expand the operators.  Compared to the present approach,
the unboundedness of the position 
and momentum operators complicates the proof of completeness. Secondly, the product rule of these 
operators is not as simple as that of the Weyl algebra. This complicates the implementation on the level of 
classical variables.  The current approach therefore seems to be more generic.

\section{Metrics from Moyal products}

For a variety of reasons it may turn out that the Hamiltonian, $H$, of a system is not hermitian with respect 
to the inner product on the Hilbert space under considerations.  This was realized some time ago in the context 
of the bosonization of fermionic systems \cite{geyer,Doba} and more recently in the context of PT-symmetric quantum 
mechanics \cite{bender}.  The important point to realize is that although the Hamiltonian may not be hermitian, it 
is not necessarily unphysical.  Indeed in the case of bosonization full equivalence with the hermitian fermion 
problem follows from construction by (non-unitary) similarity transformation, including reality of the spectrum. 
Similarly all PT-symmetric Hamilatonians have a real spectrum when PT-symmetry is not spontaneously broken 
\cite{bender}. 
The central question is then
whether a consistent quantum mechanical interpretation remains possible.  This was 
answered in \cite{scholtz} where it was pointed out that a normal quantum mechanical interpretation is possible 
if a metric operator $\Theta$ exists which has as domain the whole Hilbert space, is hermitian, positive definite and 
bounded, and satisfies the equation
\begin{equation}
\label{metricdef}
H\Theta=\Theta H^\dagger\,.
\end{equation}
Once the existence of such an operator has been established, a new inner product can be defined with respect 
to which the Hamiltonian is hermitian and a standard quantum mechanical interpretation is possible.  However, 
as was pointed out in \cite{scholtz}, also discussed in \cite{most1,most2}, the condition (\ref{metricdef}) is not 
sufficient to fix the metric uniquely, which implies that the quantum mechanical interpretation based on this 
metric, and associated inner product, is ambiguous.  The metric is uniquely determined (up to an irrelevant global 
normalization) if one requires hermiticity of a complete irreducible set of observables, $A_i$ (of which the 
Hamiltonian may be a member), with respect to the inner product associated with $\Theta$, i.e., it is required that 
(\ref{metricdef}) holds for all observables \cite{scholtz}:
\begin{equation}
\label{metricdefg}
A_i\Theta=\Theta A_i^\dagger\,\,\forall i\,.
\end{equation}
From this point of view, the choice of observables determines the metric and Hilbert space of the quantum 
system uniquely.

One may take an alternative point of view by arguing that if a metric, satisfying (\ref{metricdef}), can be found, 
a particular choice of metric determines the allowed set of measurable observables. This is the spirit of 
PT-symmetric quantum mechanics.  

Irrespective of the particular point of view, for applications it is important to be able to solve the defining 
equation (\ref{metricdefg}) in order to investigate its implications and the role of the non-uniqueness.  As 
this is a highly non-trivial problem, partially addressed in \cite{scholtz}, several recent papers addressed 
this issue from several viewpoints \cite{most2,swanson,bender}.  Here we present a new approach to this problem, 
based on the Moyal product, in which (\ref{metricdefg}) is no longer an operator equation, but is transformed into 
a differential equation for the function $\Theta(x,p)$, as defined in (\ref{functionq}). 

Let us consider the defining equation (\ref{metricdefg}) on the level of the Moyal product formulation.  On this 
level the observables $A_i$, their hermitian conjugates $A_i^\dagger$ and the metric operator $\Theta$ get replaced 
by functions $A_i(x,p)$, $A_i^\dagger (x,p)$ and $\Theta(x,p)$ as prescribed in (\ref{functionq}).  
Note that $A_i(x,p)$ 
and $A_i^\dagger (x,p)$ are related as in (\ref{hcq}).  In terms of these functions the defining relation 
(\ref{metricdefg}) reads
\begin{equation}
\label{metricdefgm}
A_i(x,p)\ast \Theta(x,p)=\Theta(x,p)\ast A_i^\dagger(x,p)\,.
\end{equation}        

We start our analysis of this equation by considering the simplest case where we require that $\hat x$ and 
$\hat p$ are hermitian observables.  In this case the equation (\ref{metricdefgm}) simply becomes 
$\Theta^{(1,0)}(x,p)=\Theta^{(0,1)}(x,p)=0$, i.e., the metric is just a constant.  This simply reflects the following 
facts: (1) that $\hat x$ and $\hat p$ form an irreducible set, which implies that the metric is uniquely 
determined up to a global normalization factor and (2) that we have chosen $\hat x$ and $\hat p$ hermitian 
from the outset (insisting on unitary representations of the Heisenberg-Weyl algebra), so that it must be 
proportional to the identity (the original inner product corresponds to $\Theta=1$). 

To proceed it is convenient to choose a particular model.  For this purpose we take the PT-symmetric model 
with Hamiltonian \cite{bender}
\begin{equation}
\label{ham}
H(\hat x,\hat p)=\hat p^2 +ig\hat x^3\,.
\end{equation}
Here we consider the Hamiltonian to be one of the observables and leave the other possible observables 
unspecified.  We expect that this may lead to non-uniqueness of the metric.  This may, however, be informative 
as the precise origin of this non-uniqueness may provide an alternative viewpoint to the existing one that 
requires the specification of a complete set of observables in order to fix the metric uniquely. On the level 
of the Moyal product this Hamiltonian gets replaced by the function (it is a sum of functions depending on 
$\hat p$ and $\hat x$ only)
\begin{equation}
\label{hamf}
H(x,p)=p^2 +igx^3\,,
\end{equation}
while the metric becomes a function $\Theta(x,p)$ as defined in (\ref{functionq}).  For the Hamiltonian 
equation (\ref{metricdefg}) now reads
\begin{equation}
\label{metricdefm}
H(x,p)\ast \Theta(x,p)=\Theta(x,p)\ast H^\ast(x,p)\,.
\end{equation}
Note that in this case the hermitian conjugate of the operator gets replaced by the complex conjugate of the 
function $H(x,p)$, i.e., $H^\dagger(x,p)=H^\ast(x,p)$.  It is simple to see that this is a generic feature of 
functions (or the sum of functions) that depend on $\hat x$ or $\hat p$ only, as there is no phase due to the 
exchange of the operators $e^{it\hat p}$ and $e^{is\hat x}$ (see (\ref{ponly})). 

Before proceeding a few comments on (\ref{metricdefm}) and its relation to Wigner functions  are in order\footnote{We are indebted to the referee for drawing our attention to this point}.  When the Hamiltonian is hermitian, which, for Hamiltonians of the form $H=p^2+V(x)$, implies that the function $H(x,p)$ is real, equation (\ref{metricdefm}) is the condition for a stationary Wigner function \cite{moyal,wigner,curtright} and $\Theta(x,p)$ indeed coincides with the Wigner function of a mixed state with equal weights assigned to the pure states as we take $\Theta=1$ in this case \cite{wigner}. Let $P_E$ denote the projection operator on an eigenstate of the Hamiltonian with eigenvalue $E$.  Projecting the metric operator $\Theta$ onto this eigenstate, i.e., $\Theta_E\equiv P_E\Theta P_E$, it follows from (\ref{metricdef}) that $[H,\Theta_E]=0$ and $H\Theta_E=\Theta_E H=E\Theta_E$.  On the level of the Moyal product this reads $H(x,p)\ast\Theta_E(x,p)=\Theta_E(x,p)\ast H(x,p)$ and $H(x,p)\ast\Theta_E(x,p)=\Theta_E(x,p)\ast H(x,p)=E\Theta_E(x,p)$.  The latter is the `star-genvalue' equation of \cite{curtright}.  In this case $\Theta_E$ corresponds to the Wigner function of a pure state.  In the hermitian case, therefore, the metric corresponds to the Wigner function of a specific mixed state, while the projected metric corresponds to the Wigner function of a pure state.  In this regard, note that since the pure state Wigner function corresponds to the projected metric, it cannot conform to the defining properties (positive definiteness) of the metric, which necessarily corresponds to the Wigner function of a mixed state with all pure state probabilities non-zero.  For this reason, from a metric point of view, one does not study the `star-genvalue' equation, which is only satisfied by pure state Wigner functions, but just the stationarity condition (\ref{metricdefm}) which also applies to mixed state Wigner functions. 

In the non-hermitian case, not considered previously from the Wigner function point of view, there is a strong analogy between the metric analysis here and Wigner functions, but there are also crucial differences, both on the technical and conceptual levels, that need to be pointed out. Firstly,  the equation (\ref{metricdefm}) is no longer a commutator equation and can not simply be interpreted as a stationarity condition.  Secondly, the 'star-genvalue' equation has to be generalized, through the introduction of appropriate left and right projection operators,  to $H(x,p)\ast\Theta_E(x,p)=E\Theta_E(x,p)$ and $\Theta_E(x,p)\ast H^\dagger(x,p)=E^*\Theta_E(x,p)$, taking into account that the eigenvalues may now be in general complex. Thirdly, it has to be noted that in the hermitian case the momentum integral of the Wigner function provides a probability distribution.  Here this is no longer necessarily the case, i.e., this integral may be negative or even complex.  This is precisely the aim of the present approach, namely, to establish the existence and construction of a new inner product which does admit a normal quantum mechanical interpretation, given that the Hamiltonian and other observables are non-hermitian.  Fourthly, the metric is actually defined with respect to a complete set of observables (see (\ref{metricdefgm})), which is conceptually quite distinct from the Wigner function approach.  Finally, as in the hermitian case, the metric must be, from a Wigner function point of view, related to a mixed state with all pure state probabilities non-zero.  For this reason one has to focus on equation (\ref{metricdefm}) and not the corresponding 'star-genvalue' equations, applicable to pure states only.                 

Returning to equation (\ref{metricdefm}), we note that since $H(x,p)$ is polynomial, 
the Moyal product truncates and one obtains the following differential equation for $\Theta(x,p)$:
\begin{equation}
\label{diffeq}
2\,i \,g\,x^3\,\Theta(x,p) - 3\,g\,\hbar\,x^2\,\Theta^{(0,1)}(x,p) - 
  3\,i \,g\,\hbar^2\,x\,\Theta^{(0,2)}(x,p) + g\,\hbar^3\,\Theta^{(0,3)}(x,p) - 
  2\,i \,\hbar\,p\,\Theta^{(1,0)}(x,p) + \hbar^2\,\Theta^{(2,0)}(x,p)=0\,.
\end{equation}
To simplify the equation, the notation $\Theta^{(n,m)}=\frac{\partial^{n+m}\Theta}{\partial^n x\partial^m p}$ has been introduced.  This is an exact equation, valid to all orders in the coupling strength $g$ and $\hbar$.  Note that since no 
boundary conditions have been specified, the solution is not unique. On the level of the Moyal products the 
non-uniqueness of the metric therefore resides in the freedom to specify the boundary conditions in (\ref{diffeq}).  
It should, however, be noted that the boundary condition that is to be imposed on (\ref{diffeq}) can not be 
arbitrary as the solution has to conform with the conditions of hermiticity and positive definiteness of the 
metric.  Keeping in mind that the metric is uniquely specified once a complete set of observables, hermitian with 
respect to the inner product associated with $\Theta$,  is identified, suggests an interplay between the boundary 
conditions imposed on (\ref{diffeq}) and the choice of physical observables on the operator level. In this regard, 
note that a choice of boundary conditions that does not admit a solution conforming to hermiticity and positive 
definitess, constitute an inconsistent choice of observables and subsequently an inconsitent quantum system as 
was pointed out in \cite{scholtz}. On the other hand, if an appropriate choice of boundary conditions, which 
fixes the metric uniquely, is made, both the Hilbert space and allowed observables in the quantum theory are 
uniquely determined.  The allowed observables can indeed be computed by solving (\ref{metricdefgm}) for the 
observables once the metric has been determined.  In this case each choice of boundary condition corresponds to 
an admissable observable.

The first issue that can be addressed directly from this equation is the existence of an hermitian metric operator, 
$\Theta$, as required by the definition of the metric operator.  Equation (\ref{diffeq}) is clearly linear and of the 
form $L\Theta(x,p)=0$, with $L$ a differential operator.  Using $e^{-i\hbar \partial_x\partial_p}x
e^{i\hbar \partial_x\partial_p}=x-i\hbar\partial_p$ and $e^{-i\hbar \partial_x\partial_p}p
e^{i\hbar \partial_x\partial_p}=p-i\hbar\partial_x$, one easily verifies $e^{-i\hbar \partial_x\partial_p}L
e^{i\hbar \partial_x\partial_p}= -L^*$, so that $L^*e^{-i\hbar \partial_x\partial_p}\Theta(x,p)=0$.  On the other 
hand the complex conjugate of (\ref{diffeq}) reads $L^*\Theta^*(x,p)=0$. Provided that the boundary condition 
imposed on (\ref{diffeq}) also satisfy (\ref{ccq}), uniqueness of the solution ((\ref{diffeq}) is linear) 
implies $\Theta^*(x,p)=e^{-i\hbar \partial_x\partial_p}\Theta(x,p)$.  Thus, provided that the boundary conditions 
imposed are consistent with (\ref{ccq}), the solution of (\ref{diffeq}), when employed to construct the metric 
operator as described in the previous section, will yield an hermitian metric operator.

Before proceeding to discuss particular solutions of (\ref{diffeq}) it is useful to consider the general 
properties such solutions must have.  The first property we note is that if $\Theta(x,p)$ is a solution 
(not necessarily corresponding to an hermitian and positive definite operator) of (\ref{diffeq}), 
or equivalently (\ref{metricdefm}), then for arbitrary functions $f(H(x,p))$ and $g(H^\dagger(x,p))$ 
the following is also a solution: $f(H(x,p))\ast \Theta(x,p)\ast g(H^\dagger(x,p))$, where the functions $f$ 
and $g$ are defined through a Taylor expansion involving the Moyal product.  This is most easily checked 
directly on the level of equation (\ref{metricdefm}) using the associativity of the Moyal product and the 
fact that $f(H)\ast H=H\ast f(H)$ and $g(H^\dagger)\ast H^\dagger=H^\dagger\ast g(H^\dagger)$. This is 
merely a reflection of the non-uniqueness of the solution of (\ref{diffeq}), which has to be eliminated 
through some appropriate choice of boundary conditions.  As was pointed out earlier the boundary conditions 
can not be completely arbitrary, but it must conform with hermiticity and positive definiteness.  This does, 
however, not eliminate the freedom in (\ref{diffeq}) completely and more input is required in the form of 
boundary conditions. Indeed, one can easily verify from (\ref{metricdefm}), associativity and the relation 
(\ref{hcq}) that if $\Theta(x,p)$ is a solution corresponding to an hermitian and positive definite operator, 
so will $g(H(x,p))\ast \Theta(x,p)\ast e^{i\hbar\partial_x\partial_p}g(H(x,p))^\ast$.

Using the above properties it is in fact possible to construct a very general solution to (\ref{diffeq}).  
The first point to note is that $e^{2ixp/\hbar}$ solves (\ref{diffeq}), although it does not satisfy the 
criterion for hermiticity.  As above one can now construct a wide class of solutions from this one, also 
not generally satisfying the criterion of hermiticity. It is in fact possible to construct a solution which 
does meet this criterion. To do this note that $e^{-2ixp/\hbar}$ solves the complex conjugate 
of (\ref{diffeq}).  Using the property $e^{i\hbar \partial_x\partial_p}L^\ast e^{-i\hbar \partial_x\partial_p}=-L$, 
one easily verifies that $e^{i\hbar \partial_x\partial_p}e^{-2ixp/\hbar}$ solves (\ref{diffeq}).  Due to 
the linearity of (\ref{diffeq}) it then follows that 
$a e^{2ixp/\hbar}+a^\ast e^{i\hbar \partial_x\partial_p}e^{-2ixp/\hbar}$, where $a$ is an arbitrary 
constant, is a solution which also satisfies (\ref{ccq}), i.e., it corresponds to a hermitian operator.  
This method of constructing a solution conforming to hermiticity is indeed very generic. As above a wide 
class of solutions, satisfying the criterion of hermiticity, can now be constructed similarly.   
However, these solutions do not have the required asymptotic behaviour when $g$ approaches zero 
(where they are supposed 
to reduce to a constant -- see the discussion below (\ref{metricdefgm})) and we do not consider them further.   

Solutions with the correct asymptotic behaviour can be constructed by resorting to a series representation in 
$g$ for the solutions of (\ref{diffeq}).  This also allows us to make contact with existing literature in 
which a series expansion was used \cite{most2}.  Once the solution has been found, albeit as a series, 
the criterion (\ref{ccq}) can be used to test for hermiticity and positive definiteness.

We list the result to $O(g)$ below, as the expression becomes rather elaborate at higher order, but 
in principle it is quite simple to compute the result to any order desired:
\begin{eqnarray}
\label{solog}
&&\Theta(x,p)=\frac{-21\,i \,e^{\frac{2\,i \,p\,x}{\hbar}}\,g\,\hbar^4\,c1(p)}{16p^6} - 
  \frac{i \,e^{\frac{2\,i \,p\,x}{\hbar}}\,\hbar\,c1(p)}{2p} - 
  \frac{21\,e^{\frac{ 2\,i  \,p\,x}{\hbar}}\,g\,\hbar^3\,x\,c1(p)}{8\,p^5} + 
  \frac{i\,e^{\frac{2\,i \,p\,x}{\hbar}}\,g\,\hbar^2\,x^2\,c1(p)}{8p^4} + \nonumber\\
 && \frac{e^{\frac{ 2\,i  \,p\,x}{\hbar}}\,g\,\hbar\,x^3\,c1(p)}{4\,p^3} + c2(p) + 
  \frac{3\,i \,g\,\hbar^2\,x\,c2(p)}{4p^4} - \frac{3\,g\,\hbar\,x^2\,c2(p)}{4\,p^3} - 
  \frac{i\,g\,x^3\,c2(p)}{2p^2} + \frac{g\,x^4\,c2(p)}{4\,\hbar\,p} - 
  \frac{i\,e^{\frac{2\,i \,p\,x}{\hbar}}\,g\,\hbar\,c3(p)}{2p} + \nonumber\\
  && g\,c4(p) + \frac{21\, i\,e^{\frac{2\,i  \,p\,x}{\hbar}}\,g\,\hbar^4\,
     c1'(p)}{16p^5} + \frac{21\,e^{\frac{ 2\,i  \,p\,x}{\hbar}}\,g\,\hbar^3\,x\,c1'(p)}
   {8\,p^4} - \frac{9\,i\,e^{\frac{2\,i  \,p\,x}{\hbar}}\,g\,\hbar^2\,x^2\,c1'(p)}
   {8p^3} - \frac{e^{\frac{2\,i \,p\,x}{\hbar}}\,g\,\hbar\,x^3\,c1'(p)}{4\,p^2} - \nonumber\\
  &&\frac{3\,i\,g\,\hbar^2\,x\,c2'(p)}{4p^3} + \frac{3\,g\,\hbar\,x^2\,c2'(p)}{4\,p^2} + 
  \frac{i\,g\,x^3\,c2'(p)}{2p} - 
  \frac{9\,i \,e^{\frac{2\,i \,p\,x}{\hbar}}\,g\,\hbar^4\,c1''(p)}{16p^4} - 
  \frac{9\,e^{\frac{2\,i \,p\,x}{\hbar}}\,g\,\hbar^3\,x\,c1''(p)}{8\,p^3} + \nonumber\\
  &&\frac{\,i\,e^{\frac{2\,i \,p\,x}{\hbar}}\,g\,\hbar^2\,x^2\,c1''(p)}{8p^2} + 
 \frac{3\,i \,g\,\hbar^2\,x\,c2''(p)}{4p^2} - \frac{3\,g\,\hbar\,x^2\,c2''(p)}{4\,p} + 
 \frac{i \,e^{\frac{2\,i  \,p\,x}{\hbar}}\,g\,\hbar^4\,c1^{(3)}(p)}{8p^3} + \nonumber\\
  &&\frac{e^{\frac{ 2\,i  \,p\,x}{\hbar}}\,g\,\hbar^3\,x\,c1^{(3)}(p)}{4\,p^2} - 
  \frac{i\,g\,\hbar^2\,x\,c2^{(3)}(p)}{2p}\,+O(g^2)\,.
\end{eqnarray}
This is the most general form of the solution where $c1(p)$, $c2(p)$, $c3(p)$ and $c4(p)$ are completely 
arbitrary functions of $p$. These functions are, however, restricted if one requires that the expansion (\ref{solog}) 
satisfies the hermiticity condition (\ref{ccq}).  Indeed, one can quite easily see that $c2(p)$ and $c4(p)$ 
must at least be real.  
Here we do not pursue the most general solution, but rather focus on a particular choice of integration 
constants for which this expression simplifies considerably. Setting $c1(p)=0$, $c2(p)=1$ (this brings about 
only a global normalization), $c3(p)=0$ and $c4(p)=0$ one finds to $O(g)$
\begin{equation}
\label{solbcg}
\Theta(x,p)=1 + g\,\left( \frac{3\,i\,\hbar^2\,x}{4p^4} - \frac{3\,\hbar\,x^2}{4\,p^3} - \frac{i \,x^3}{2p^2} + 
     \frac{x^4}{4\,\hbar\,p} \right)\,.
\end{equation}
This result agrees with the result in \cite{most2} when the different normalization of the kinetic energy  term, 
accounting for the factor of $\frac{1}{2}$, and the different convention for the definitions of the metric 
(see (\ref{metricdef})), which brings about a complex conjugation, are taken into account.  With this choice of 
boundary conditions, even the higher order term simplifies considerably and one easily finds to $O(g^3)$ 
(setting any further integration constants zero)   

\begin{eqnarray}
\label{solbcg2}
\Theta(x,p)&=&1 + g\,\left( \frac{3\,i \,\hbar^2\,x}{4p^4} - \frac{3\,\hbar\,x^2}{4\,p^3} - \frac{i\,x^3}{2p^2} + 
     \frac{x^4}{4\,\hbar\,p} \right) \nonumber\\
     & +& g^2\,\left( \frac{108\,i  \,\hbar^5\,x}{p^9} - \frac{108\,\hbar^4\,x^2}{p^8} - 
     \frac{ 57\,i  \,\hbar^3\,x^3}{p^7} + \frac{21\,\hbar^2\,x^4}{p^6} + 
     \frac{6\,i  \,\hbar\,x^5}{p^5} - \frac{11\,x^6}{8\,p^4} - \frac{i \,x^7}{4\hbar\,p^3} + 
     \frac{x^8}{32\,\hbar^2\,p^2} \right)  \nonumber\\
     &+& g^3\,\left( \frac{29872557\,i \,\hbar^8\,x}{256p^{14}} - 
     \frac{29872557\,\hbar^7\,x^2}{256\,p^{13}} - \frac{7676559\,i\,\hbar^6\,x^3}{128p^{12}} + 
     \frac{5395599\,\hbar^5\,x^4}{256\,p^{11}} + \frac{727299\,i\,\hbar^4\,x^5}{128p^{10}}\right.\nonumber\\
     & -& \left.\frac{159489\,\hbar^3\,x^6}{128\,p^9}
     -\frac{14679\,i \,\hbar^2\,x^7}{64p^8} + \frac{9207\,\hbar\,x^8}{256\,p^7} + 
     \frac{615\,i\,x^9}{128p^6} - \frac{343\,x^{10}}{640\,\hbar\,p^5} - 
     \frac{3\,i \,x^{11}}{64\hbar^2\,p^4} + \frac{x^{12}}{384\,\hbar^3\,p^3} \right)\,+O(g^4)\nonumber\\
     &\equiv& 1+ga+g^2b+g^3c+O(g^4)\,.
\end{eqnarray}   
Note that $\Theta(x,p)$ is singular in $\hbar$.  Indeed, one easily establishes that this is an essential 
singularity 
(of the type $e^{a/\hbar}$), so that, not surprisingly, the metric does not exist at the classical level.  

Next we consider the issue of positive definiteness of $\Theta$.  For this one has to show that the logarithm of 
$\Theta$ 
is 
also hermitian.  Let us compute the logartihm of $\Theta$ from (\ref{solbcg2}).  To do this, we must keep in mind 
that 
in order to reflect the operator character correctly, the logarithm must be computed using Moyal products.  
One easily finds to $O(g^3)$
\begin{equation}
\label{logT1}
(\log \Theta)(x,p)=\log(1+ga+g^2b+g^3c)=ag+g^2\left(b-\frac{a\ast a}{2}\right)+g^3\left(c+\frac{a\ast a\ast a}{3}-
\frac{a\ast b+b\ast a}{2}\right)\,.
\end{equation}
Due the polynomial nature of the functions $a$, $b$ and $c$ in $x$, the Moyal products truncate and can be 
evaluated easily to yield
\begin{eqnarray}
\label{logT2}
(\log \Theta)(x,p)&=&g\,\left( \frac{3\,i\,\hbar^2\,x}{4p^4} - \frac{3\,\hbar\,x^2}{4\,p^3} - \frac{i \,x^3}{2p^2} + 
     \frac{x^4}{4\,\hbar\,p} \right)\nonumber\\
     &+&g^3\left(\frac{-2745171\,i\,\hbar^8\,x}{256p^{14}} + \frac{2745171\,\hbar^7\,x^2}{256\,p^{13}} + 
  \frac{677457\,i\,\hbar^6\,x^3}{128p^{12}} - \frac{439857\,\hbar^5\,x^4}{256\,p^{11}} - 
  \frac{52029\,i\,\hbar^4\,x^5}{128p^{10}} \right.\nonumber\\
  &+&\left. \frac{9375\,\hbar^3\,x^6}{128\,p^9} + 
  \frac{651\,i\,\hbar^2\,x^7}{64p^8} - \frac{273\,\hbar\,x^8}{256\,p^7} - \frac{5\,i \,x^9}{64p^6} + 
  \frac{x^{10}}{320\,\hbar\,p^5}\right)\,.
\end{eqnarray}
Note that the second order term vanishes with this choice of boundary conditions \cite{most2}.

Finally, we check the hermiticity of $\log \Theta$, i.e., we have to verify if (\ref{ccq}) holds.  Clearly this must 
happen order by order in $g$.  It is easily verified from (\ref{logT2}) that this is indeed the case up to 
$O(g^3)$.  We have now verified, at least on the level of the series expansion, that both the metric and its 
logaritm is hermitian, implying that $\Theta$ is also positive definite.  

As a final example we consider the following Hamiltonian, which has often been discussed in the context of
non-hermitian Hamiltonians.   In second quantized form it reads
\begin{equation}
\label{ham1}
H=\hbar\omega a^\dagger a+\hbar\alpha aa+\hbar\beta a^\dagger a^\dagger\,.
\end{equation}
A finite dimensional version of this model was studied in \cite{scholtz} and more recently this model was studied 
in \cite{swanson, geyer1,jones}. This can be rewritten in terms of $\hat x$ and $\hat p$ in the usual way by setting 
$a^\dagger=(\hat x-i\hat p)/\sqrt{2\hbar}$ and $a=(\hat x+i\hat p)/\sqrt{2\hbar}$.  Suppressing irrelevant 
constant terms, this yields
\begin{equation}
H=a\hat p^2+b\hat x^2+ic\hat p\hat x\,,\;a=(\omega-\alpha-\beta)/2,\,b=(\omega+\alpha+\beta)/2,\,c
=(\alpha-\beta)\,.
\end{equation}
On the level of the Moyal bracket formulation this becomes
\begin{equation}
\label{clasham1}
H(x,p)=a p^2+b x^2+icpx;\quad H^\dagger(x,p)=a p^2+b x^2-icpx+c\hbar\,.
\end{equation}
Substituting this in (\ref{metricdefm}) yields the equation for the metric
\begin{eqnarray}
\label{diffeq1}
&&c\left(\hbar - 2\,i\,p\,x \right) \,\Theta(x,p) \nonumber\\
&&+\hbar\,\left( \left( c\,p - 2\,i\,b\,x \right) \,\Theta^{(0,1)}(x,p)+ 
     \left( c\,x+2\,i \,a\,p\,\right)\Theta^{(1,0)}(x,p) + b\,\hbar\,\Theta^{(0,2)}(x,p)  - 
a\,\hbar\,\Theta^{(2,0)}(x,p) 
\right)=0\,.
\end{eqnarray}

As before one easily verifies that the solution of this equation satisfies (\ref{ccq}), provided that 
appropriate boundary conditions are imposed, which confirms that the metric operator constructed from this 
solution is hermitian.     

In this case one can find an exact solution to (\ref{diffeq1}) as a one parameter family of metrics, given by
\begin{equation}
\label{onepar}
\Theta(x,p)=e^{r\,p^2 + s\,p\,x + t\,x^2},\,
\end{equation}
where
\begin{equation}
\label{onepar1}
r=\frac{-c \pm {\sqrt{c^2 - 4\,a\,b\,\hbar\,s\left(2i-\hbar s\right)}}}{4\,b\,\hbar}\,,t
=\frac{c \pm {\sqrt{c^2 - 4\,a\,b\,\hbar\,s\left(2i-\hbar s\right)}}}{4\,a\,\hbar}\,,
\end{equation}
$s$ being a free parameter.  Once this solution has been found, a large class of solutions can be constructed as was pointed out earlier.  

For an arbitrary choice of $s$ the solution (\ref{onepar}) does not have the required asymptotic behaviour in the limit of a hermitian Hamiltonian, i.e., when $c=0$.  In this limit the metric should coincide with that of the original inner-product for which $\Theta=1$. The correct asymptotic behaviour can be obtained with appropriate choices of $s$.  As a first example, consider the choice $s=0$. With this choice $r=0$ and $t=\frac{c}{2a\hbar}$ or $r=-\frac{c}{2b\hbar}$ and $t=0$, depending on the solution taken in (\ref{onepar1}). Since $a$, $b$ and $c$ are real, the condition (\ref{ccq}) is trivially satisfied in both these cases and the metric is hermitian. To show positive definiteness one has to verify that the logarithm of the metric is hermitian. In the Moyal product formulation this implies that one has to find the function corresponding to the logarithm of the metric operator and verify that it satisfies (\ref{ccq}), i.e., one has to find the function $\eta(x,p)$ such that $1+\eta+\frac{1}{2!}\eta\ast\eta+\frac{1}{3!}\eta\ast\eta\ast\eta\ast+\ldots=\Theta$.  In this case it is, however, obvious that the Moyal product reduces to an ordinary product so that the function corresponding to the logarithm of the metric operator is simply the logarithm of (\ref{onepar}), which is $-\,c\,p^2/2b\hbar$ or $\,c\,x^2/2a\hbar$, depending on the choice of solution in (\ref{onepar1}).  This trivially satisfies (\ref{ccq}) so that the metric is positive definite, although not necessarily bounded.      

The choice $s=0$ made above is but one of an infinite family that yields the appropriate asymptotic behaviour of the metric when $c=0$.  Indeed, it is easily verified that the following choice, where $\epsilon^2<1$, also satisfies this requirement: $s=\frac{i}{\hbar}(1-\sqrt{1+\frac{\epsilon^2 c^2}{4|ab|}})$, yielding $r=\frac{-c\pm \sqrt{c^2(1-\epsilon^2{\rm sgn}(ab))}}{4b\hbar}$ and $t=\frac{c\pm \sqrt{c^2(1-\epsilon^2{\rm sgn}(ab))}}{4a\hbar}$.  The hermiticity and positive definiteness of this metric can now be verified from a series expansion of this solution in $c$.  As the technical steps are identical to the computation in the previous example, we do not repeat them here.

\section{Conclusions}

We have shown how the Moyal product can be used to compute the metric for a given non-hermitian Hamiltonian.  
The verification that the metric is hermitian and positive definite can be carried out directly on the level 
of the Moyal product formulation, without refering to the operator level. We have carried through this program 
for two Hamiltonians, both of which possess PT-symmetry.  These considerations can also be applied to finite 
dimensional models as those studied in \cite{scholtz}, by using the finite dimensional formulation of the Moyal 
product, where, essentially, $h\rightarrow\frac{1}{N}$.  An interesting new perspective that arises from 
this formulation is the relation between the choice of observables and boundary conditions imposed on the 
metric equation, as formulated in terms of the Moyal product, both of which give rise to a unique metric.  

Finally we mention that these considerations generalize to many types of operator equations such as 
(\ref{metricdef}).  Other examples that will be discussed elsewhere involve the computation of the 
Berry connection \cite{scholtz1} and the implementation of Wegner's flow equations \cite{scholtz2}.

\section*{Acknowledgements}

The authors acknowledge financial support from the National Research Foundation of South Africa.


\begin{thebibliography}{99}
\bibitem{bbj} C M Bender, D C Brody and H F Jones, Am. J. Phys. 71 (2003) 1095.
\bibitem{bender} C M Bender, Contemp. Phys. 46 (2005) 277.
\bibitem{scholtz} F G Scholtz, H B Geyer and F J W Hahne, Ann. Phys. 213 (1992)74.
\bibitem{most1} A Mostafazadeh, J. Math. Phys. 43 (2002) 205; {\it ibid} 2814; {\it ibid} 3944.
\bibitem{most2} A Mostafazadeh, quant-ph/0508195.
\bibitem{moyal} J E Moyal, Proc. Camb. Phil. Soc. 45 (1949) 99.
\bibitem{fairlie} D Fairlie, Mod. Phys. Lett. A 13 (1998) 263.
\bibitem{fairlie1} D B Fairlie, P Fletcher and C K Zachos, Phys. Lett. B 218 (1989) 203.
\bibitem{bratteli} O Bratteli and D W Robinson, {\it Operator Algebras and Quantum Statistical Mechanics}, 
Vol.1, Springer, Heidelberg, 1987.
\bibitem{bender1} C M Bender and G V Dunne, Phys. Rev. D 40 (1988) 2739; {\it ibid} (1989) 3504.
\bibitem{geyer}   H B Geyer and F J W Hahne, Nucl. Phys. A 363 (1981) 45.
\bibitem{Doba} J Dobaczewski, Nucl. Phys. A 369 (1981) 213;
               {\it ibid} 217; {\it ibid} A 380 (1982) 1.
\bibitem{swanson} M Swanson, J. Math. Phys. 45 (2004) 585.  
\bibitem{geyer1} H B Geyer, F G Scholtz and I Snyman, Czech. Jnl. Phys. 54 (2004) 1069.
\bibitem{jones} H F Jones, J. Phys. A 38 (2005) 1741.
\bibitem{wigner} E Wigner, Phys. Rev. 40 (1932) 749.
\bibitem{curtright} T Curtright, D Fairlie and C Zachos, Phys. Rev. D 58 (1998) 025002.
\bibitem{scholtz1} F G Scholtz and H B Geyer, in preperation.
\bibitem{scholtz2} J N Kriel, F G Scholtz and J D Thom, in preparation.
\end{thebibliography}
\end{document}